\documentclass[final,5p,times,twocolumn]{elsarticle}

\usepackage{graphicx}
\usepackage{amssymb}
\usepackage{lipsum}
\usepackage{mathtools}
\usepackage{hyperref}
\usepackage{makecell}
\usepackage{subcaption}
\usepackage{graphicx}
\usepackage{caption}
\usepackage{multirow}
\usepackage[all]{xy}
\usepackage{booktabs}
\usepackage[skip=0.1\baselineskip]{caption}

\usepackage{amsmath}

\usepackage{natbib}
\biboptions{semicolon,round}

\setcitestyle{authoryear,open={(},close={)}}

\hypersetup{
     colorlinks   = true,
     citecolor    = blue
}

\journal{NeuroImage: Clinical}

\begin{document}

\begin{frontmatter}


\title{Cross-Cohort Generalizability of Deep and Conventional Machine Learning for MRI-based Diagnosis and Prediction of Alzheimer’s Disease}

\author[1]{Esther E. Bron}
\ead{e.bron@erasmusmc.nl}
\author[1]{Stefan Klein}
\author[2]{Janne M. Papma}
\author[2]{Lize C. Jiskoot}
\author[1]{Vikram Venkatraghavan}
\author[1]{Jara Linders}
\author[3]{Pauline Aalten}
\author[4]{Peter Paul De Deyn}
\author[5]{Geert Jan Biessels}
\author[7]{Jurgen A.H.R. Claassen}
\author[8,11]{Huub A.M. Middelkoop}
\author[1]{Marion Smits}
\author[1,9]{Wiro J. Niessen}
\author[2]{John C. van Swieten}
\author[10]{Wiesje M. van der Flier}
\author[3]{Inez H.G.B. Ramakers}
\author[1]{Aad van der Lugt}
\author{for the Alzheimer's Disease Neuroimaging Initiative\fnref{fn1}}
\fntext[fn1]{Data used in preparation of this article were obtained from the Alzheimer’s Disease Neuroimaging Initiative (ADNI) database (\url{adni.loni.usc.edu}). As such, the investigators within the ADNI contributed to the design and implementation of ADNI and/or provided data
but did not participate in analysis or writing of this report. A complete listing of ADNI investigators can be found at: \url{http://adni.loni.usc.edu/wp-content/uploads/how\_to\_apply/ADNI\_Acknowledgement\_List.pdf}}
\author{on behalf of the Parelsnoer Neurodegenerative Diseases study group\fnref{fn2}}
\fntext[fn2]{Data used in preparation of this article were obtained from the Health-RI Parelsnoer Neurodegenerative Diseases Biobank (\url{www.health-ri.nl/parelsnoer}).} 

\address[1]{Department of Radiology and Nuclear Medicine, Erasmus MC, Rotterdam, the Netherlands}
\address[2]{Department of Neurology, Erasmus MC, Rotterdam, the Netherlands}
\address[3]{Alzheimer Center Limburg, School for Mental Health and Neuroscience (MHeNS), Maastricht University Medical Center, Maastricht, The Netherlands}
\address[4]{Department of Neurology and Alzheimer Center, University Medical Center Groningen, Groningen, The Netherlands}
\address[5]{Department of Neurology, UMC Utrecht Brain Center, University Medical Center Utrecht, Utrecht, The Netherlands}
\address[7]{Radboud University Medical Center, Nijmegen, The Netherlands}
\address[8]{Department of Neurology \& Neuropsychology, Leiden University Medical Center, Leiden, The Netherlands}
\address[11]{Institute of Psychology, Health, Medical and Neuropsychology Unit, Leiden University, The Netherlands}
\address[9]{Imaging Physics, Applied Sciences, Delft University of Technology, the Netherlands}
\address[10]{Amsterdam University Medical Center, location VUmc, Amsterdam, The Netherlands}

\begin{abstract}
This work validates the generalizability of MRI-based classification of Alzheimer's disease (AD) patients and controls (CN) to an external data set and to the task of prediction of conversion to AD in individuals with mild cognitive impairment (MCI). 

We used a conventional support vector machine (SVM) and a deep convolutional neural network (CNN) approach based on structural MRI scans that underwent either minimal pre-processing or more extensive pre-processing into modulated gray matter (GM) maps. Classifiers were optimized and evaluated using cross-validation in the Alzheimer's Disease Neuroimaging Initiative (ADNI; 334 AD, 520 CN). Trained classifiers were subsequently applied to predict conversion to AD in ADNI MCI patients (231 converters, 628 non-converters) and in the independent Health-RI Parelsnoer Neurodegenerative Diseases Biobank data set. From this multi-center study representing a tertiary memory clinic population, we included 199 AD patients, 139 participants with subjective cognitive decline, 48 MCI patients converting to dementia, and 91 MCI patients who did not convert to dementia. 

AD-CN classification based on modulated GM maps resulted in a similar area-under-the-curve (AUC) for SVM (0.940; 95\%CI: 0.924-0.955) and CNN (0.933; 95\%CI: 0.918-0.948). Application to conversion prediction in MCI yielded significantly higher performance for SVM (AUC=0.756; $95\%$CI: $0.720-0.788$) than for CNN (AUC=0.742; $95\%$CI: $0.709-0.776$) ($p<0.01$ for McNemar’s test). In external validation, performance was slightly decreased. For AD-CN, it again gave similar AUCs for SVM (0.896; 95\%CI: 0.855-0.932) and CNN (0.876; 95\%CI: 0.836-0.913). For prediction in MCI, performances decreased for both SVM (AUC=0.665; $95\%CI$: $0.576-0.760$) and CNN (AUC=0.702; $95\%$CI: $0.624-0.786$). Both with SVM and CNN, classification based on modulated GM maps significantly outperformed classification based on minimally processed images ($p=0.01)$. 

Deep and conventional classifiers performed equally well for AD classification and their performance decreased only slightly when applied to the external cohort. We expect that this work on external validation contributes towards translation of machine learning to clinical practice.

\end{abstract}


\begin{keyword}
Alzheimer's disease \sep Support vector machine \sep Convolutional Neural Network \sep External validation


\end{keyword}

\end{frontmatter}



\newcommand\blfootnote[1]{%
  \begingroup
  \renewcommand\thefootnote{}\footnote{#1}%
  \addtocounter{footnote}{-1}%
  \endgroup
}

\DeclareRobustCommand{\de}[3]{#3}

\newcommand{\beginsupplement}{%
        \setcounter{table}{0}
        \renewcommand{\thetable}{S\arabic{table}}%
        \setcounter{figure}{0}
        \renewcommand{\thefigure}{S\arabic{figure}}%
     }

\section{Introduction}

The diagnostic process of dementia is challenging and takes a substantial period of time after the first clinical symptoms arise: on average 2.8 years in late-onset and 4.4 years in young-onset dementia \citep{vanvliet2013}. The window of opportunity for advancing the diagnostic process is however much larger than these few years. For Alzheimer's disease (AD), the most common form of dementia, there is increasing evidence that disease processes start 20 years or more ahead of clinical symptoms \citep{Gordon2018}. Advancing the diagnosis is essential to support the development of new disease modifying treatments, since late treatment is expected to be a major factor in the failure of clinical trials \citep{Mehta2017}. In addition, early and accurate diagnosis have great potential to reduce healthcare costs as they give patients access to supportive therapies that help to delay institutionalization \citep{prince2011alzheimer}. 

Machine learning offers an approach for automatic classification by learning complex and subtle patterns from high-dimensional data. In AD research, such algorithms have been frequently developed to perform automatic diagnosis and predict the future clinical status at an individual level based on biomarkers. These algorithms aim to facilitate medical decision support by providing a potentially more objective diagnosis than that obtained by conventional clinical criteria  \citep{Kloppel2012, Rathore2017}. A large body of research has been published on classification of AD and its prodromal stage, mild cognitive impairment (MCI) \citep{Ansart2021, Wen2020, Rathore2017, Arbabshirani2017, Falahati2014, Bron2015}. Overall, classification methods show high performance for classification of AD patients and control participants with an area under the receiver-operating characteristic curve (AUC) of 85-98\%. Reported performances are somewhat lower for prediction of conversion to AD in patients with MCI (AUC: 62-82\%). Structural T1-weighted (T1w) MRI to quantify neuronal loss is the most commonly used biomarker, whereas the support vector machine (SVM) is the most commonly used classifier. Following the trends and successes in medical image analysis and machine learning, neural network classifiers - convolutional neural networks (CNN) in particular - have increasingly been used since few years \citep{Wen2020, Cui2019, Basaia2019}, but have not been shown to significantly outperform conventional classifiers. Most CNN studies perform no to minimal pre-processing of the structural MRI scans as input for their classifier \citep{Wen2020, Basaia2019, Hosseini-Asl2018, Vieira2017}, while others use more extensive pre-processing strategies proven successful for conventional classifiers, such as gray matter (GM) density maps \citep{Cui2019, Suk2017}. Although CNNs are designed to extract high-level features from raw imaging data, it is imaginable that the learning process for complex tasks is improved by dedicated pre-processing that enhances disease-related features, which reduces model complexity and enables a more stable learning process. It is unclear yet whether CNNs would improve AD classification over conventional classifiers and whether they benefit from extensive MRI pre-processing.




Despite high performance of machine learning diagnosis and prediction methods for AD, it is largely unknown how these algorithms would perform in clinical practice. A next step would be to assess the generalizability of classification methods from a specific research population to another study population. \textcolor{black}{There are however only very few studies assessing classification performance on an external data set \citep{Wen2020,Bouts2019,Archetti2019,Hall2015}. Results varied from only a minor reduction in performance for some experiments \citep{Wen2020, Hall2015} to a severe drop for others \citep{Bouts2019,Archetti2019,Wen2020}.} While generalizability seemed related to how well the training data represented the testing data (e.g. an external data set with similar inclusion criteria showed a smaller performance drop than a data set with very different criteria \citep{Wen2020}, a better understanding is crucial before applying such methods in routine clinical practice. 



Therefore, this work aims to assess the generalizability of MRI-based classification performance to an external data set representing a tertiary memory clinic population for both diagnosis of AD and prediction of AD in individuals with MCI. To evaluate the value of neural networks and to determine their optimal MRI pre-processing approach, we compare a CNN with a conventional SVM classifier using two pre-processing approaches: minimal pre-processing using only rough spatial alignment and more extensive pre-processing into modulated GM maps. First, we optimize the methods using a large research cohort and assess classification performance using cross-validation. Subsequently, we validate AD prediction performance in MCI patients of the same cohort as well as AD diagnosis and prediction performance in the external data set.

\begin{table}[bt]
\caption{Demographics for the ADNI data set. \label{tab:adni-demographics}}
\begin{center}
\begin{footnotesize}
\begin{tabular}{ c c c c c } 
\toprule
 & AD & CN & MCIc & MCInc  \\
\cmidrule(lr){2-3}
\cmidrule(lr){4-5}
\emph{\# participants} & 336 & 520 & 231 & 628 \\
\emph{male / female} & 186 / 150 & 252 / 268 & 141 / 90 &  367 / 261 \\
\emph{age (y; mean$\pm$std)} & $74.9\pm7.8$  & $74.2\pm5.8$ & $73.7\pm7.0$ & $72.9\pm7.8$ \\ 
 \bottomrule
\end{tabular}
\end{footnotesize}
\end{center}
\end{table}

\begin{table}[bt]
\caption{Demographics for the PND data set. FU: follow-up time \label{tab:pnd-demographics}}
\begin{center}
\begin{footnotesize}
\begin{tabular}{ c c c c c } 
\toprule
 & AD & SCD & MCIc & MCInc  \\
\cmidrule(lr){2-3}
\cmidrule(lr){4-5}
\emph{\# participants} & 199 & 138 & 48 & 91 \\
\emph{male / female} & 94 / 105 & 92 / 46 & 33 / 15 &  56 / 35 \\
\emph{age (y; mean$\pm$std)} & $73.1\pm9.6$  & $63.2\pm10.3$ & $70.4\pm7.9$ & $68.8\pm12.6$ \\ 
\emph{FU (y; mean$\pm$std)} & N.A. & N.A. & $2.7 \pm 1.2$ & $2.2 \pm 0.8$\\
 \bottomrule
\end{tabular}
\end{footnotesize}
\end{center}
\end{table}

\begin{table*}[!htb]
\caption{An overview of T1-weighted imaging protocols in the PND data set from eight centers. All sequences were 3D and used gradient recalled echo (GRE). TFE: turbo field echo, FSPGR: fast spoiled GRE, TFL: turboflash, MPRAGE: magnetization prepared rapid gradient echo, MP: magnetization prepared, SS: steady state, SP: spoiled, IR: inversion recovery, Sag: sagittal, Cor: Coronal, Ax: axial, TE: echo time, TR: repetition time, TI: inversion time. \label{tab:pnd-imaging}}
\begin{center}
\begin{scriptsize} 
\begin{tabular}{ ccccccccccccccc} 
\toprule
\emph{Field strength} & \multicolumn{6}{c}{3T} & \multicolumn{5}{c
}{1.5T} \\
\cmidrule(lr){1-1}
\cmidrule(lr){2-7}
\cmidrule(lr){8-12}
\emph{Vendor} & \multicolumn{4}{c}{Philips} & GE & Siemens & Philips & \multicolumn{2}{c}{GE} & \multicolumn{2}{c}{Siemens}\\
\cmidrule(lr){1-1}
\cmidrule(lr){2-5}
\cmidrule(lr){6-6}
\cmidrule(lr){7-7}
\cmidrule(lr){8-8}
\cmidrule(lr){9-10}
\cmidrule(lr){11-12}
\emph{Number of scans} & 195 & 60 & 70 & 1 & 122 & 40 & 1 & 6 & 1 & 32 & 28\\
\emph{Sequence name} & TFE & GRE & GRE & GRE & FSPGR & MPRAGE & GRE & FSPGR & FSPGR & TFL MPRAGE & MPRAGE \\
\emph{Sequence variant} & MP & MP & MP & MP & SS/SP & IR/SP/MP & MP & SS/SP & SS/SP & IR/SP/MP & IR/SP/MP \\
\emph{Plane} & Sag & Sag & Sag & Cor & Sag & Sag & Sag & Sag & Ax & Sag & Sag
\\\
\emph{Slice thickness (mm)} & 1 & 1 & 1 & 1.4 & 1 & 1 & 2 & 1.6 & 1.6 & 1 & 1\\
\emph{In-plane (mm*mm)} & 1*1 & 0.78*0.78 & 0.5*0.5 & 0.88*0.88  & 0.94*0.94 & 1*1 & 0.47*0.47 & 0.47*0.47 & 1*1 & 1*1 & 0.5*0.5\\
\emph{TE (ms)} & 4.2  & 4.6  & 3.5 & 4.6 & 3 & 4.7 & 4.6 & 2.1 & 4.2 & 3 & 3.7 \\
\emph{TR (ms)} & 8.1 & 9.9 & 9 & 9.6 & 7.8 & 2300 & 15 & 7.1 & 8.3 & 2000 & 2700 \\
\emph{TI (ms)} & N.A. & N.A. & N.A. & N.A. & 450 & 1100 & N.A. & 450 & 450 & 1100 & 950 \\
\bottomrule
\end{tabular}
\end{scriptsize}
\end{center}
\end{table*}

\section{Methods}
\subsection{Study population}
We used data from two cohorts. The first group of 1715 participants was included from the Alzheimer's Disease Neuroimaging Initiative (ADNI; \url{adni.loni.usc.edu}). The ADNI was launched in 2003 as a public-private partnership, led by Principal Investigator Michael W. Weiner, MD. The primary goal of ADNI has been to test whether clinical and neuropsychological assessment, serial magnetic resonance imaging (MRI), positron emission tomography (PET), and other biological markers can be combined to measure the progression of mild cognitive impairment (MCI) and early Alzheimer’s disease (AD). For up-to-date information, see \url{www.adni-info.org}. We included all participants with a T1w MRI scan at baseline from the ADNI1/GO/2 cohorts: 336 AD patients, 520 control participants (CN), 231 mild cognitive impaired (MCI) patients who converted to AD within 3 years (MCIc) and 628 MCI patients who did not convert (MCInc). The CN group consisted of 414 cognitively normal participants and 106 participants with subjective cognitive decline (SCD). Demographics are shown in Table \ref{tab:adni-demographics}. A list of included participants is made available at \url{https://gitlab.com/radiology/neuro/bron-cross-cohort}. 

The second group of participants was included from the Health-RI Parelsnoer Neurodegenerative Diseases Biobank (PND; \url{www.health-ri.nl/parelsnoer}), a collaborative biobanking initiative of the eight university medical centers in the Netherlands \citep{Mannien2017}. The Parelsnoer Neurodegenerative Diseases Biobank focuses on the role of biomarkers on diagnosis and the course of neurodegenerative diseases, in particular of Alzheimer’s disease \citep{Aalten2014}. It is a prospective, multi-center cohort study, focusing on tertiary memory clinic patients with cognitive problems including dementia. Patients are enrolled from March 2009 and followed annually for two to five years. In the PND biobank, a total of 1026 participants have been included. Inclusion criteria for the current research were: a high resolution T1w MRI at baseline, clinical consult at baseline, 90 days or less between MRI and clinical consult, and a baseline diagnosis of SCD, MCI, or dementia due to AD. A flow diagram of the inclusion can be found in the supplementary files (Fig. \ref{fig:pnd-inclusion}). A total of 557 participants met inclusion criteria. One person was excluded because image analysis failed. This led to inclusion of 199 AD patients and 138 participants with SCD. Of the MCI group, we included the 139 participants that had a follow-up period of at least 6 months. Of this group, 48 MCI patients converted towards dementia within the available follow-up time and 91 MCI patients remained stable. Demographics are shown in Table \ref{tab:pnd-demographics}.

\begin{figure*}[!htb]
\centering
   \includegraphics[width=1\linewidth]{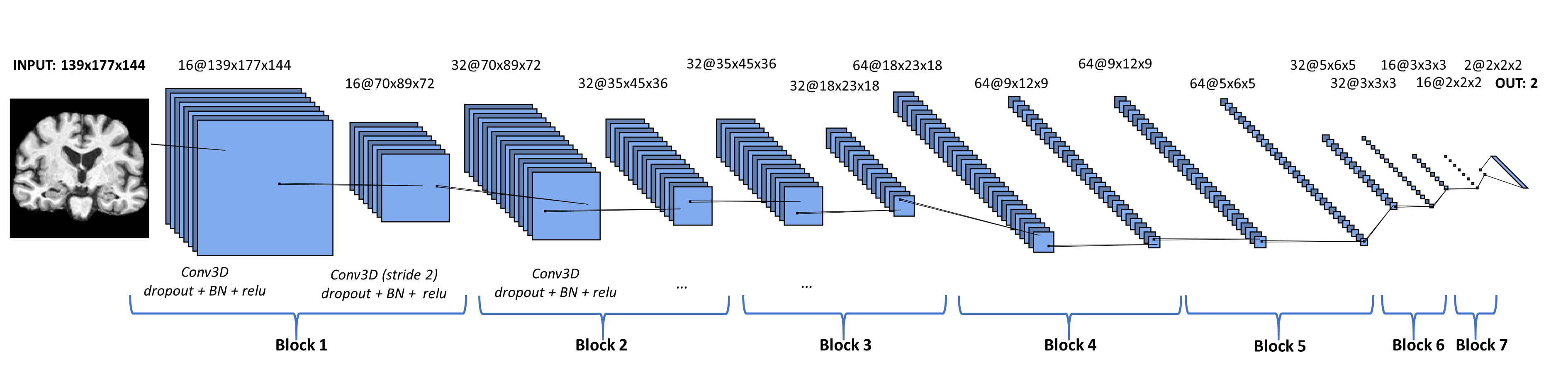}
\caption[cnn]{CNN architecture\footnotemark}
\label{fig:cnn}
\end{figure*}

\subsection{Imaging data}
We used baseline T1w structural MRI acquired at 1.5T or 3T. Acquisition protocols are previously described (ADNI: \citet{Jack2008, Jack2015}, PND: \citet{Aalten2014}). Variation in acquisition protocols used in PND is detailed in Table \ref{tab:pnd-imaging}. For the majority of scans, a 8-channel head coil was used (N=423; 76\%); other scans used a 16-channel (N=27), 24-channel (N=1), 40-channel (N=1), or unknown coil (N=104). 



\subsection{Image pre-processing}
We evaluated two pre-processing approaches based on T1w images: minimal pre-preprocessing and a more extensive pre-processing into modulated GM maps. 

To prepare T1w images with minimal pre-processing, scans were non-uniformity corrected using the N4 algorithm \citep{tustison2010n4itk} and subsequently transformed to MNI-space using registration of brain masks with a similarity transformation. A similarity transformation is a rigid transformation including isotropic scaling. Registrations were performed with Elastix registration software \citep{klein2010elastix, Shamonin2014}. To account for variations in signal intensity, images were normalized within the brain mask to have zero mean and unit variance. 

To obtain modulated GM maps encoding gray matter density, the Iris pipeline was used \citep{Bron2014HBM}. To compute these maps a group template space was defined using a procedure that avoids bias towards any of the individual T1w images using pairwise registration \citep{seghers2004construction}. The pairwise registrations were performed using a similarity, affine, and nonrigid B-spline transformation model consecutively. We selected a subset of images for the definition of the template space. This template set consisted of the images of 50 ADNI participants that were randomly selected preserving the ratio between diagnostic groups (subject list available at \url{https://gitlab.com/radiology/neuro/bron-cross-cohort}). The other images of both ADNI and PND data sets were registered to the template space following the same registration procedure. For the current work, some changes to the template space construction procedure as used in \cite{Bron2014HBM} were made: non-uniformity correction was performed, skull-stripping was performed, and the template space corresponded to MNI-space. Using similarity registration based on brain masks, we computed the coordinate transformations of MNI space to each of the template set's images, which were subsequently concatenated with the pairwise transformations before averaging. After template space construction, probabilistic GM maps were obtained with the unified tissue segmentation method of SPM8 (Statistical Parametric Mapping) \citep{ashburner2005tissueseg}. To obtain the final feature maps, probabilistic GM maps were transformed to the template space and modulated, i.e. multiplied by the Jacobian determinant of the deformation field, to take compression and expansion into account \citep{ashburner2000voxel}. To correct for head size, modulated GM maps were divided by intracranial volume.

\subsection{Classification approaches}
Two machine learning approaches were used for classification: a support vector machine (SVM) and a convolutional neural network (CNN). 

\subsubsection{Support vector machine (SVM)}
An SVM with a linear kernel was used as this approach previously showed good performance using voxel-based features for AD classification. \citep{kloppel2008automatic, cuingnet2011automatic, Bron2014HBM, Bron2015}. The c-parameter was optimized with 5-fold cross-validation on the training set. Input features, i.e. voxel values of the pre-processed images within a brain mask, were normalized to zero mean and unit variance based on the training set. The classifier was implemented using Scikit-Learn. 

To gain insight into the classifications, we  calculated statistical significance maps (p-maps) that show which features contributed to the SVM decision. These maps were computed using an analytical expression that approximates permutation testing \citep{Gaonkar2015}. Clusters of significant voxels were obtained using a p-value threshold of $\alpha\leq 0.05$. P-maps were not corrected for multiple comparisons, as permutation testing has a low false-positive rate \citep{gaonkar2013analytic}. 

\subsubsection{Convolutional neural network (CNN)}
\footnotetext{Figure created with \url{https://alexlenail.me/NN-SVG}} \textcolor{black}{An all convolutional neural network was used \citep{Springenberg2015}, which is a fully convolutional network (FCN) architecture that uses standard convolutional layers with stride two instead of the pooling layers used in most CNNs.} This approach was chosen as it has previously shown good classification performance for AD based on structural MRI \citep{Cui2019, Basaia2019}. The used architecture is shown in Fig. \ref{fig:cnn}. Specifically, the network was built of 7 blocks consisting of a 3D convolutional layer (filter size 3; stride 1), followed by dropout, batch normalization (BN), and a rectified linear unit (ReLU) activation function, succeeded by a second 3D convolutional layer (filter size 3; stride 2), dropout, BN, and ReLU activation \citep{Cui2019, Basaia2019}\textcolor{black}. The number of filters changed over blocks: 16 filters in block 1, 32 in block 2 and 3, 64 in block 4 and 5, 32 in block 6, and 16 in block 7. The final output layer of the network was a softmax activation function, providing 2 prediction values (1 per class). The total network consisted of 577,498 parameters.


\textcolor{black}{For artificially increasing the training data set and for removing the class imbalance, data augmentation was used. The training set was augmented to 1000 samples per class based on the `mixup' approach \citep{Eatonrosen2018, Zhang2018}. Mixup is a data-agnostic augmentation approach that is not based on spatial transformations, and therefore does not degrade the spatial normalization. Augmented samples were constructed by linearly combining two randomly selected images of the same class: a fraction of $80\%$ of the first image was added to a fraction of $20\%$ of the second image.}

The network was compiled with a binary cross-entropy loss function and Adam optimizer (learning rate=0.001, epsilon=1e-8, decay=0.0). To facilitate a stable convergence, learning rate followed a step decay schedule, i.e. after each ten epochs the learning rate was divided by two. The dropout rate was set to $20\%$. Data was propagated through the network with a batch size of 4. Input images were normalized to zero mean and unit variance based on the augmented training set. \textcolor{black}{A validation set was created by randomly splitting $10\%$ of the training data which was not used for training but only for regularization by early stopping, i.e. training was stopped when the validation AUC had not increased for 20 epochs.} The model of the epoch with the highest validation AUC was selected as final model. Implementation was based on Keras and Tensorflow.

To gain insight into the classifications, we made saliency maps that show which parts of the brain contributed the most to the prediction of the CNN, i.e. which voxels lead to increase/decrease of prediction score when changed. Saliency maps were made using guided backpropagation, changing the activation function of the output layers from softmax to linear activations \citep{Springenberg2015}. Maps were averaged over correctly classified AD patients \citep{Rieke2018}. 


\begin{figure*}[!htb]
\centering
\begin{subfigure}[b]{0.45\textwidth}
   \includegraphics[width=1\linewidth]{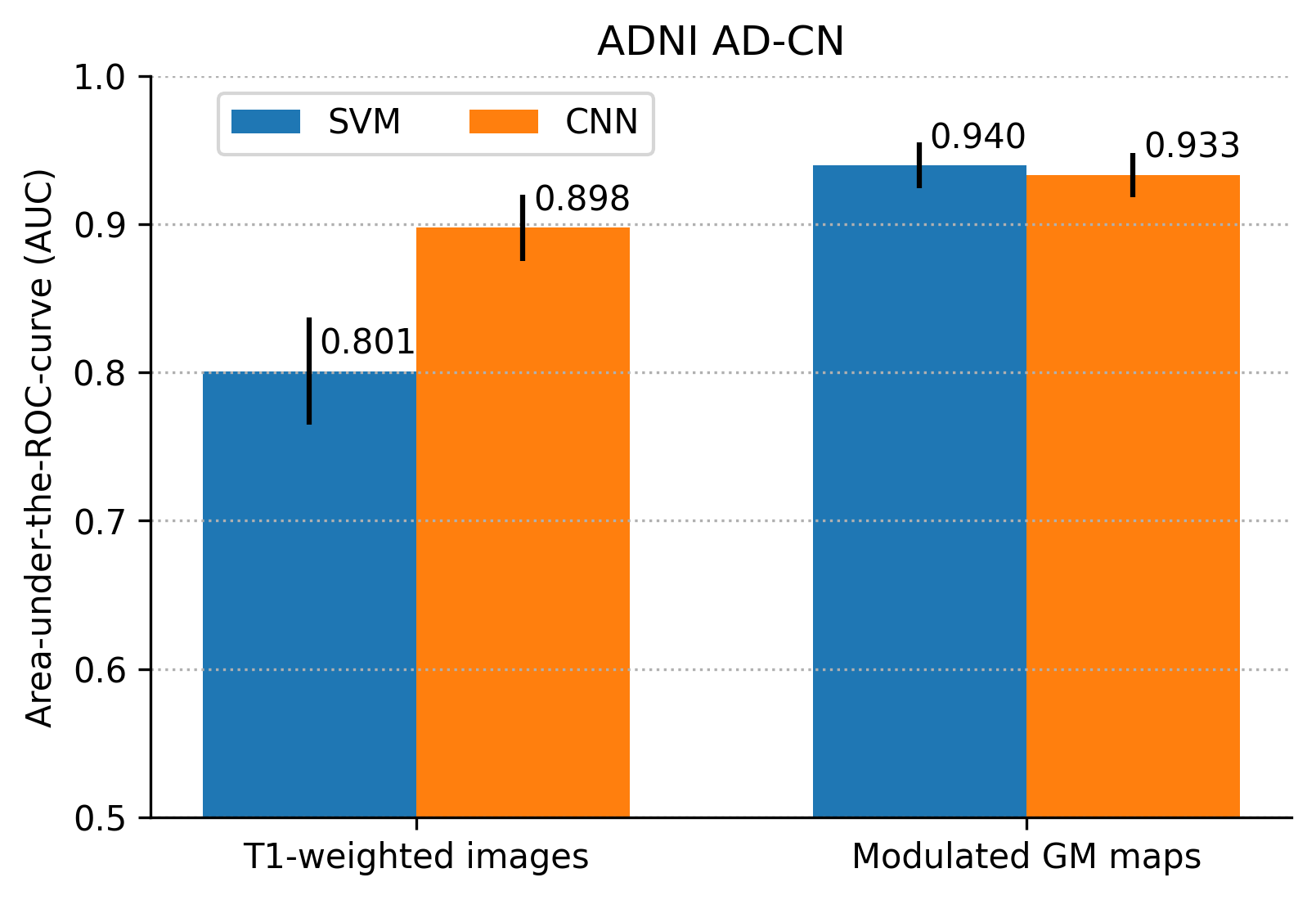}
   \caption{}
   \label{fig:adni-adcn:a} 
\end{subfigure}
~
\begin{subfigure}[b]{0.45\textwidth}
   \includegraphics[width=1\linewidth]{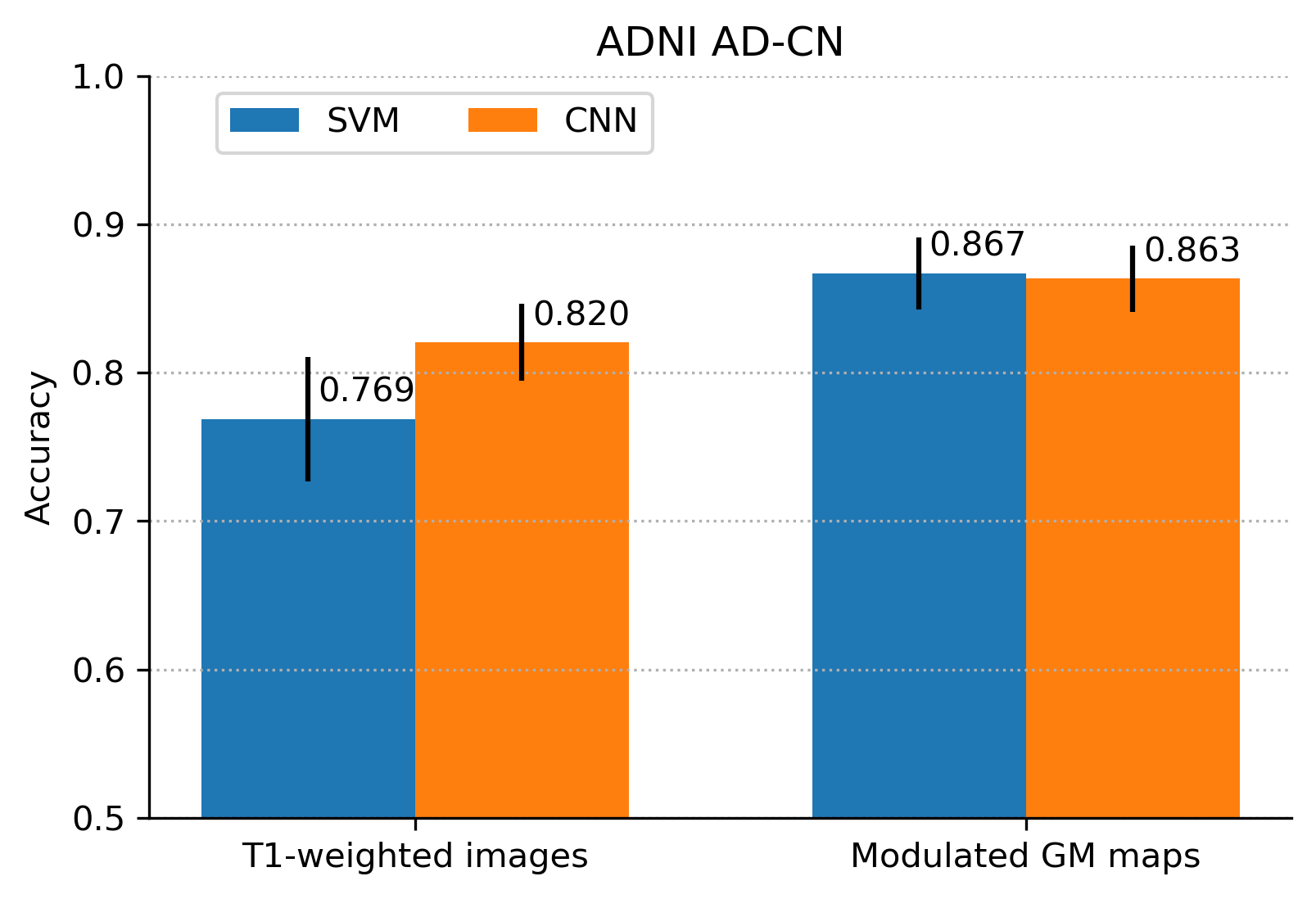}
   \caption{}
   \label{fig:adni-adcn:b}
\end{subfigure}

\caption[adni-adcn]{Cross-validation performance for classification of the Alzheimer's disease patients (AD) and controls (CN) of the ADNI data set expressed by (a) area under-the-ROC-curve (AUC) and (b) accuracy. Performance is shown for SVM and CNN classifiers using two inputs: minimally processed T1w scans and modulated GM images. Error bars indicate $95\%$CIs.}
\label{fig:adni-adcn}
\end{figure*}

\begin{figure*}[!htb]
\centering
\begin{subfigure}[b]{0.45\textwidth}
   \includegraphics[width=1\linewidth]{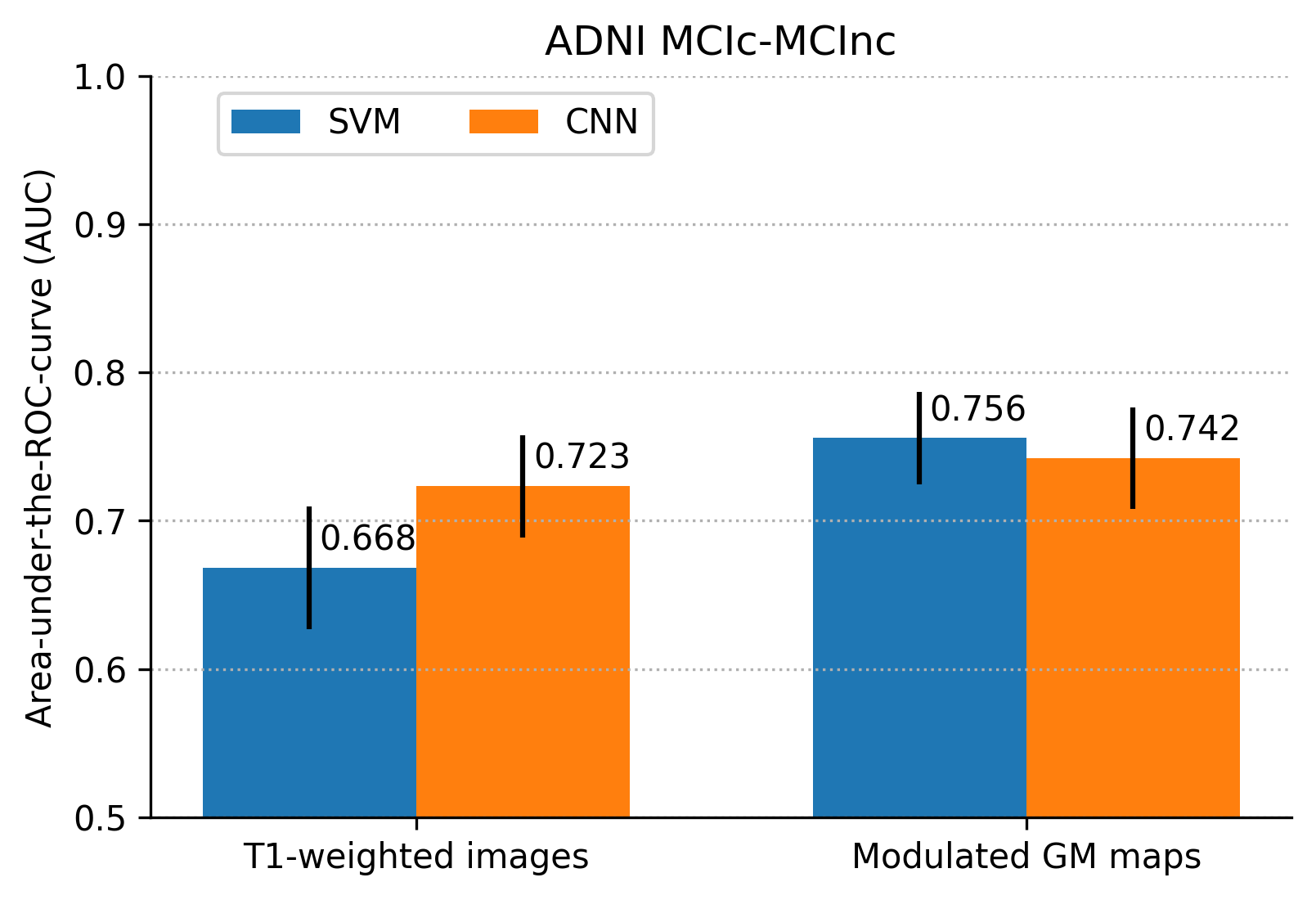}
   \caption{}
   \label{fig:adni-mci:a} 
\end{subfigure}
~
\begin{subfigure}[b]{0.45\textwidth}
   \includegraphics[width=1\linewidth]{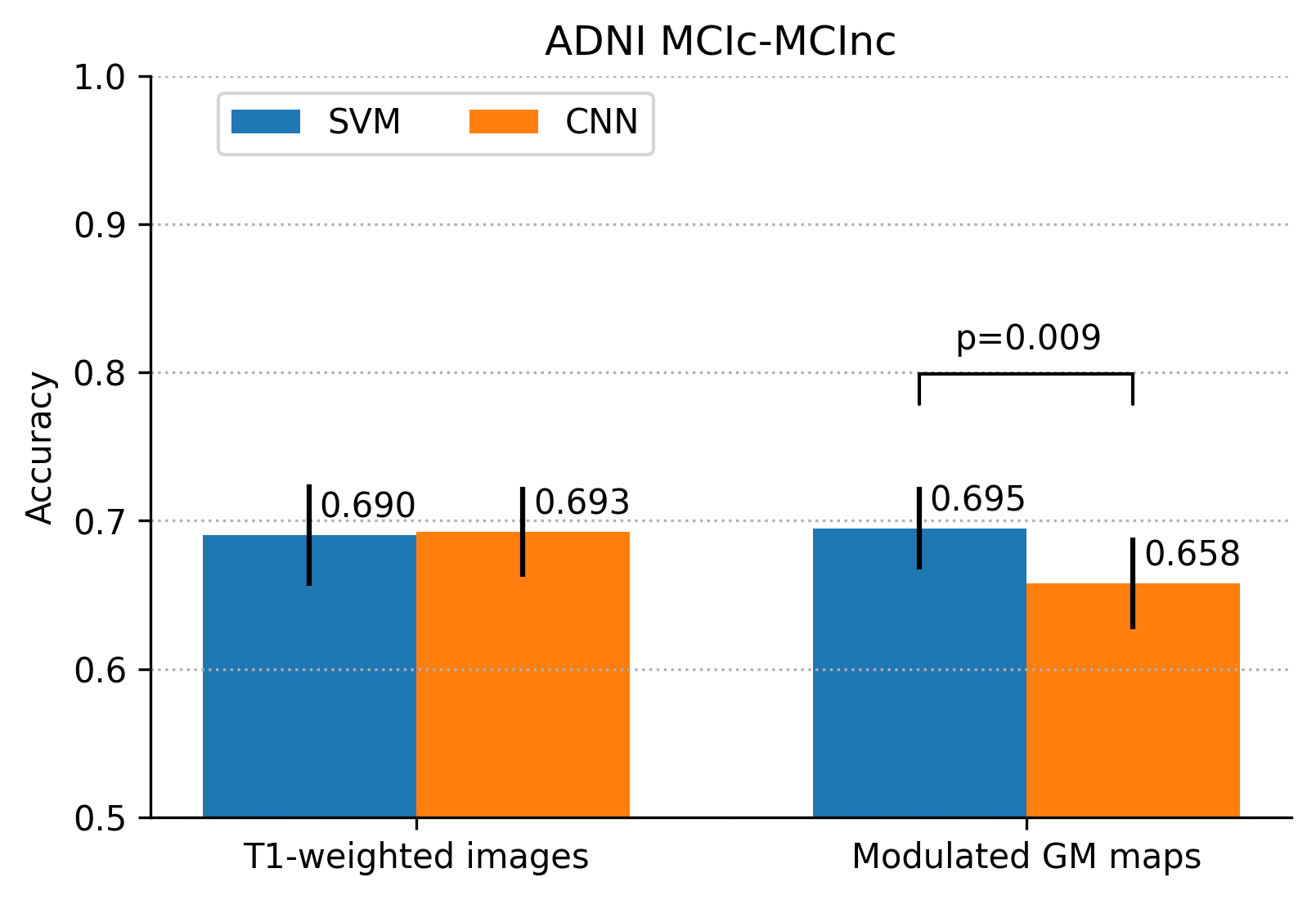}
   \caption{}
   \label{fig:adni-mci:b}
\end{subfigure}

\caption[adni-mci]{Classification performance of patients with mild cognitive impairment (MCI) that do or do not convert to Alzheimer's disease (MCIc vs MCInc) in the ADNI data set expressed by (a) area under-the-ROC-curve (AUC) and (b) accuracy. Performance is shown for SVM and CNN classifiers using two inputs: minimally processed T1w scans and modulated GM images. Error bars indicate $95\%$CIs. P-values for significant differences are shown in (b).}
\label{fig:adni-mci}
\end{figure*}

\begin{figure*}[!htb]
\centering
\begin{subfigure}[b]{0.45\textwidth}
   \includegraphics[width=1\linewidth]{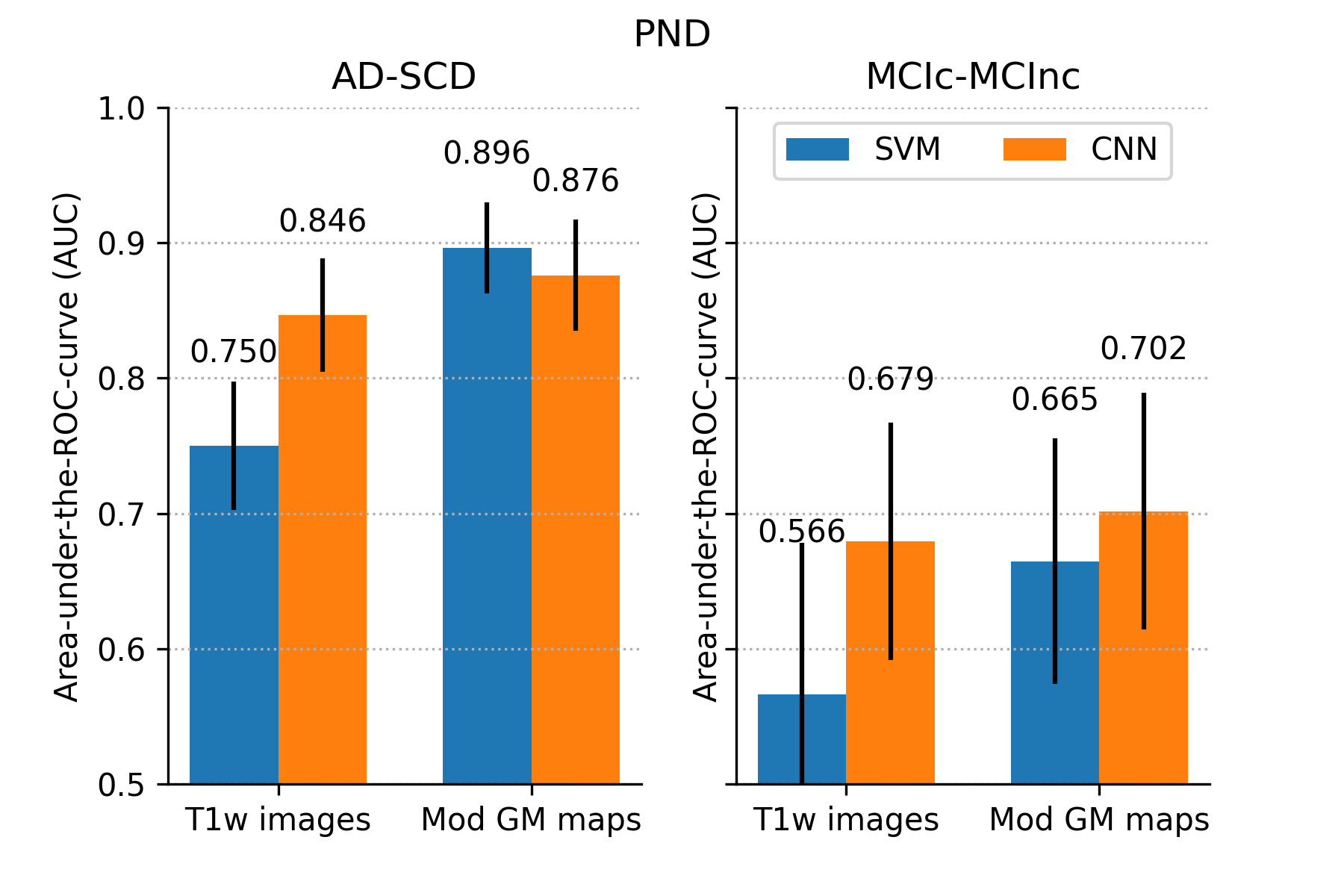}
   \caption{}
   \label{fig:pnd:a}
\end{subfigure}
~
\begin{subfigure}[b]{0.45\textwidth}
   \includegraphics[width=1\linewidth]{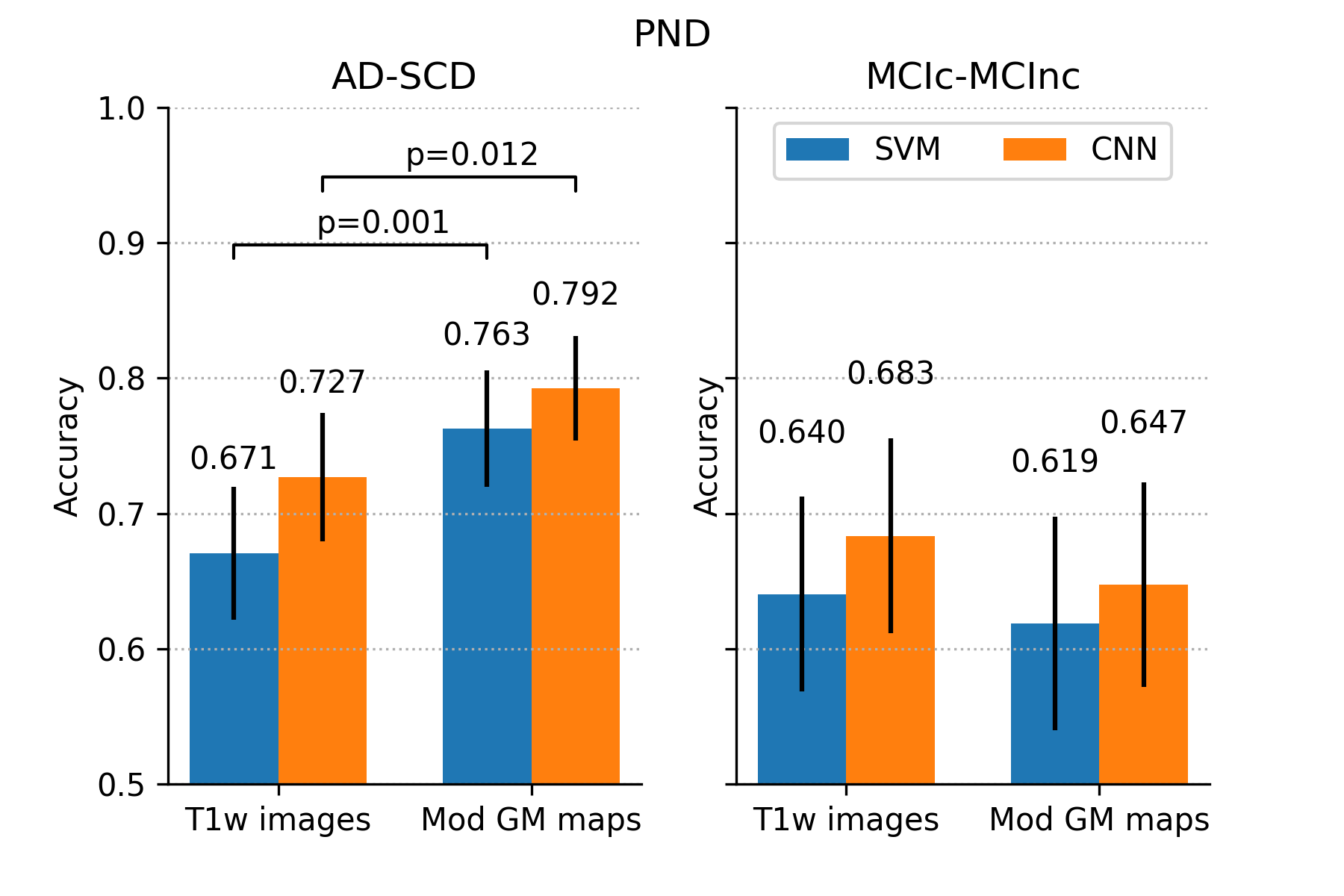}
   \caption{}
   \label{fig:pnd:b}
\end{subfigure}

\caption[pnd]{Classification performance on the PND data set: (a) area under-the-ROC-curve (AUC) and (b) accuracy. Classifiers were trained on ADNI AD-CN and applied to PND AD-SCD (left figures) and PND MCIc-MCInc (right figures). Performance is shown for SVM and CNN classifiers using two inputs: minimally processed T1w scans and modulated GM images. Error bars indicate $95\%$CIs. P-values for significant differences are shown in (b).}
\label{fig:pnd}
\end{figure*}

\subsection{Analysis and statistics}
Classification performance was quantified by the area under the curve (AUC) and accuracy. For AD-CN classification, the data of the ADNI AD and CN groups were randomly split for $20$ iterations preserving relative class sizes in each training and testing sample, using $90\%$ for training and $10\%$ for testing. Random splits were the same for both SVM and CNN. In each iteration, classification model parameters were optimized on the training set as explained above. The models were optimized solely on the training set; the test set was used only for evaluation of the final model. Ninety-five percent confidence intervals ($95\%$CI) for the mean performance measures were constructed using the corrected resampled t-test based on the 20 cross-validation iterations, thereby taking into account that the samples in the cross-validation splits were not statistically independent \citep{Nadeau}

Subsequently, we retrained classifiers using all AD and CN participants from the ADNI as training set. These retrained classifiers were used for visualization and their performance was evaluated on three independent test sets: ADNI MCIc-MCInc, PND AD-SCD, and PND MCIc-MCInc. $95\%$CIs were obtained based on 500 bootstrap samples of the test set. Significant differences between classifiers were assessed using the non-parametric McNemar Chi-square test \citep{Dietterich1998} ($\alpha<0.013$ after Bonferroni correction for 4 comparisons in each test set). 

Trained models, lists of included subjects and all code used in preparation of this article are available from \url{https://gitlab.com/radiology/neuro/bron-cross-cohort}\footnote{When this manuscript is accepted for publication, we will make a release of the repository and add a permanent DOI here}. 



\section{Results}
Cross-validation performance for the ADNI AD-CN classification is shown in Fig. \ref{fig:adni-adcn}. For SVM, the AUC using modulated GM maps ($0.940$, $95\%$CI: $0.924-0.955$) was higher than the AUC using T1w images  ($0.801$, $95\%$CI: $0.765-0.837$). For CNN, the same effect was observed, with modulated GM maps yielding a higher AUC ($0.933$, $95\%$CI: $0.918-0.948$) than T1w images ($0.898$, $95\%$CI: $0.875-0.920$), albeit here with overlapping confidence intervals. For classification based on modulated GM maps, the AUC for SVM ($0.940$; $95\%$CI: $0.924-0.955$) was similar to that of CNN ($0.933$; $95\%$CI: $0.918-0.948$). Accuracy measures showed the same patterns.

The performance of the classifiers trained on all ADNI AD and CN data to predict MCI conversion is shown in Fig. \ref{fig:adni-mci}. While AUCs with both SVM and CNN were slightly higher for modulated GM maps than for T1w images, the accuracy measures showed similar performance for both inputs. Using modulated GM maps, performance for SVM (AUC=0.756, $95\%$CI: $0.720-0.788$; accuracy=0.695, $95\%$CI: $0.665-0.723$) was higher than for CNN (AUC=0.742, $95\%$CI: $0.709-0.776$); accuracy=0.658, $95\%$CI: $0.628-0.690$). This difference was significant according to McNemar’s test ($p<0.01$). 


The performance of external validation, i.e. the application of the classifiers in the PND data set, is shown in Fig. \ref{fig:pnd}. For AD-SCD diagnosis, the AUC for SVM was $0.896$ ($95\%$CI: $0.855-0.932$) and that for CNN was $0.876$ ($95\%$CI: $0.836-0.913$). Both AUC and accuracy followed the same patterns as in ADNI: SVM and CNN showed similar performance and modulated GM maps yielded higher classification performance than minimally processed T1w images (McNemar's test; $p<0.01$ for SVM, $p=0.01$ for CNN). Performances were however slightly lower; PND confidence intervals for AUC (but not for accuracy) overlapped with those of ADNI.

For prediction of MCI conversion in PND, classification performance was also lower than that in ADNI. For the GM modulated maps, the AUC for CNN was 0.702 ($95\%$CI: $0.624-0.786$) and that for SVM was 0.665 ($95\%$CI: $0.576-0.760$). Confidence intervals were relatively large and overlapped with those in the ADNI data. No significant differences between classifiers and between pre-processing approaches were seen.

\begin{figure*}[!htb]
\centering
\includegraphics[width=0.8\linewidth]{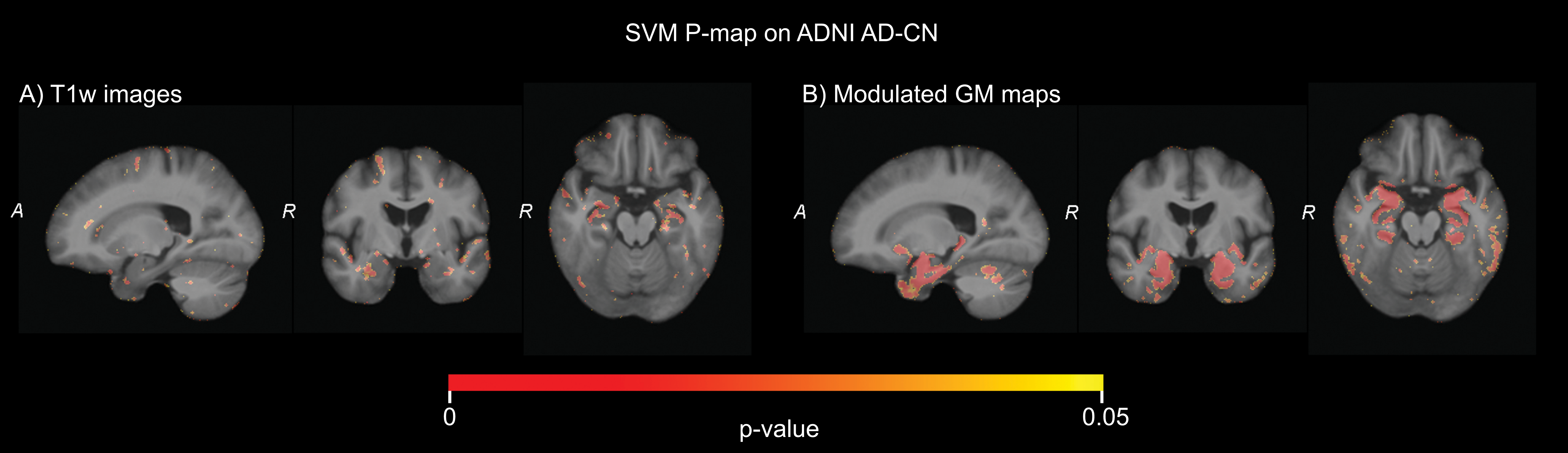}
\caption[pmap]{Visualization of the SVM classifiers using analytic significance maps (p-maps) based on two inputs: (a) minimally processed T1w images and (b) modulated GM maps.}
\label{fig:pmaps}
\end{figure*}

\begin{figure*}[!htb]
\centering
\includegraphics[width=0.8\linewidth]{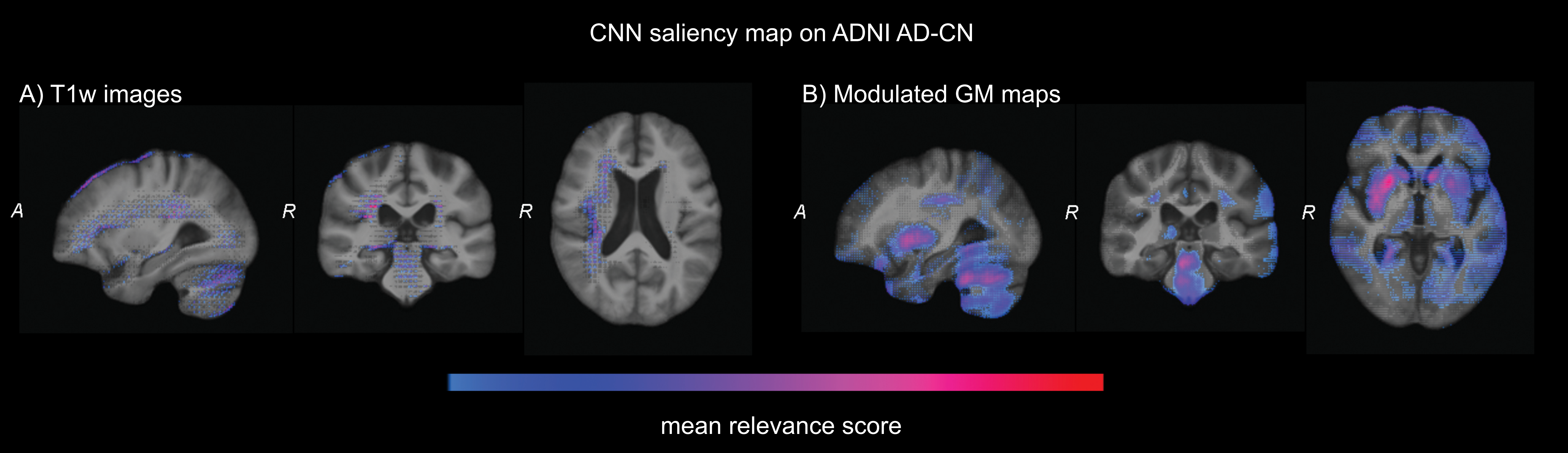}
\caption[saliency]{Visualization of the CNN classifiers using guided back-propagation saliency maps based on two inputs: (a) minimally processed T1w images and (b) modulated GM maps. Relevance maps were averaged over all correctly classified AD participants and thresholded at $\frac{1}{3}$ of the maximum intensity.}
\label{fig:saliency}
\end{figure*}

Brains regions that contributed to the classifications are visualized using SVM p-maps in Fig. \ref{fig:pmaps} and  using CNN saliency maps in Fig. \ref{fig:saliency}. The SVM p-map for the minimally processed T1w images showed small clusters of significant voxels, mainly located in the medial temporal lobe (hippocampus), around the ventricles and at larger sulci at the outside of the brain. For modulated GM maps, clusters of significant voxels in the p-map were larger and predominantly visible in the hippocampus. In addition, smaller clusters were located in the rest of the temporal lobe and the cerebellum. CNN saliency maps showed a very limited contribution of the temporal lobe. Instead, the saliency map for the T1w images mainly showed contribution of voxels at the edge of the brain, in white matter regions around the ventricles and in the cerebellum. For modulated GM maps, clusters of contributing voxels were located in the subcortical structures, the white matter around the ventricles and the cerebellum.

\section{Discussion}

We performed a comparative study focusing on the generalizability of diagnostic and predictive performance of machine learning based on MRI data of the ADNI research cohort, to the PND multi-center data set representing a tertiary memory clinic population. Both cross-validation and external validation results for AD-CN diagnosis showed similar performance using the used deep learning classifier and conventional classifier. Both approaches significantly benefited from the use of modulated GM maps instead of raw T1w images. Application to MCI conversion prediction yielded higher performance for SVM than for CNN in ADNI, but this was not seen in PND. Performances were in line with the state-of-the-art \citep{Rathore2017, Wen2020,Ansart2021}. \textcolor{black}{For MCI conversion prediction, \cite{Ansart2021} showed that performance of current methods converges to an AUC of about 75\% as the number of subjects increases, which aligns with our results.}

While in many medical imaging applications CNNs convincingly outperformed conventional classifiers \citep{Litjens2017}, our results showed similar performance for CNN and SVM, which confirms the findings by \cite{Wen2020}. \textcolor{black}{Other CNN designs could possibly improve on this, but we made an effort to follow the state-of-the-art for CNN design. Promising developments to further improve performance could come from changes in network architecture (e.g., successful standard architectures like InceptionNet or ResNet, adversarial training, discriminative auto-encoders) and improvements in data collection and handling (e.g., larger datasets to learn more complex models, or pretraining on other collections of brain imaging data).} \textcolor{red}{In addition, data augmentation could play a role in further improvement. While a strength of the mix-up approach is that it is data-agnostic, an augmentation approach using for example prior knowledge may have added value.}

\textcolor{black}{This work shows that the need for dedicated pre-processing is lower for CNN than for SVM, but nevertheless has an added value for the performance. While we evaluated only one implementation of the pre-processing procedure \citep{Bron2014HBM}, we expect that alternative implementations (e.g. SPM12, FSL-VBM) could have slightly changed results but would have led to the same conclusions.} \textcolor{red}{With sufficiently large datasets the need for dedicated pre-processing including spatial normalization may reduce.}


Although SVM and CNN classifiers yielded similar performance, their visualizations showed different brain regions to be involved in the classification. SVM significance maps showed a clear contribution of the hippocampus and medial temporal lobe as previously shown and expected based on prior knowledge \citep{Bron2017eurrad}. CNN saliency maps showed involvement of subcortical structures, regions prone to white matter hyperintensities and the cerebellum. For both classifiers, classification based on minimally processed T1w images showed voxels at the edge of the brain to be involved, which is expected as only similarity transformation to template space had been performed. \textcolor{black}{In addition to the brain edges, the CNN classifier, which outperformed the SVM for these minimally processed input images, also highlights regions similar to those shown by the saliency map for the modulated GM images. This may implicate that the CNNs non-linear operations, in contrast to the linear kernel of the SVM, could extract feature maps that partly resemble GM modulated maps.} The regions highlighted by the CNN saliency maps could possibly be related to AD using prior knowledge, but we will refrain from over-interpretation here. It is however unexpected that the medial temporal lobe is not covered as previously shown with CNN saliency maps on ADNI data \citep{Dyrba2020, Rieke2018}. Differences between the SVM and CNN classifiers in involved brain regions could be contributed to both the differences in the classification approaches as well as to the differences in the used visualization techniques. \textcolor{black}{If the first reason dominates, hence if the classifiers actually use different brain regions, combining classifiers into a hybrid approach would be an interesting future direction. However, for} full understanding of brain regions involved in CNN-based classification of AD, further research is required.

This work is one of the few to address how AD classification performance of MRI-based machine learning generalizes to an independent cohort \citep{Wen2020, Hall2015, Bouts2019, Archetti2019}. On the PND data, the resulting AUC values (0.896 for SVM, 0.876 for CNN) were competitive with values reported for AD-CN in the literature, but still they were 0.04-0.07 lower than those in the ADNI cross-validation experiment. The main patterns in the results corresponded between ADNI and PND data, i.e. similar performance for SVM and CNN and added value of dedicated MRI processing. For prediction in MCI, AUC values in the PND data set were 0.04-0.10 lower than those in ADNI. Overall, similar to experiments by \citet{Wen2020} and \citet{Hall2015}, we observed only a minor performance drop. This largely preserved performance could be related to the similarities between the ADNI and PND studies that include a multi-center set-up, within-study standardization of cognitive protocols, and diagnostic criteria for AD \citep{McKhann1984, McKhann2011} and MCI \citep{Petersen2004}. The performance reduction could be contributed to differences between the studies, such as the MRI protocols (all high resolution T1w, but more homogeneous within ADNI than within PND), country of origin (United States vs. the Netherlands), control population (a combination of cognitively normal and SCD vs. SCD only), MCI population (amnestic MCI only vs. a broad MCI group) and patient inclusion criteria (ADNI used hard cut-offs on cognitive scores and clinical dementia rating whereas PND did not) \citep{Petersen2010, Aalten2014}. Studies that found much worse generalizability in their experiments described larger differences in inclusion and diagnostic criteria between training data and validation data than we did  \citep{Bouts2019,  Wen2020}. 

A limitation of this study is that the diagnosis was based on clinical criteria rather than post-mortem histopathological examination. Although diagnosis was typically confirmed by follow-up, it is possible that some of the patients were misdiagnosed. \textcolor{black}{An alternative could be to use amyloid data from PET imaging or cerebrospinal fluid to classify AD pathology instead of relying on the clinical diagnosis (e.g., \cite{Son2020})}. In addition, because of the limited availability of diagnostic information at follow-up in the PND data set, its MCI data is relatively small. This is reflected by the large confidence intervals for the performance metrics in the prediction task. To maximize the number of PND MCI participants, we chose to use the last available time point for final diagnosis. As a result the time-to-prediction ranged between 1-5 years, whereas for ADNI a fixed time interval of three years is chosen. As time-to-prediction is related to predictive performance \citep{Ansart2021}, a fixed time interval would be preferred for inter-cohort performance comparison.

While the external validation performance was quite high, as expected some performance drop was observed. Therefore, research focusing on approaches to mitigate such performance drops, such as transfer learning, is highly relevant \citep{Wachinger2016}. In addition, whereas this work only exploited structural MRI, other works have shown that performance can be increased with the use of multi-modal inputs, i.e. cognitive test scores, fluid-based biomarker measurements, genetic information and other imaging modalities such as PET, diffusion MRI or perfusion MRI \citep{Bron2017eurrad, Ansart2021, Venkatraghavan2019}. While multi-modal classification would therefore be a logical and important extension, this may also lead to a decrease of  generalizability as chances of differences between studies increase with multiple modalities. 

In conclusion, classification performance of ADNI data generalized well to the multi-center PND biobank cohort representing tertiary memory clinic patients, with only a minor drop in performance. Conventional SVM classifiers and deep learning approaches using CNN showed comparable results, and both methods benefited from dedicated MRI processing using GM modulated maps. We hope that external validation results like those presented here will contribute to setting next steps towards the implementation of machine learning in clinical practice for aiding diagnosis and prediction. 

\section*{Acknowledgements}
The authors would like to thank Judith Manniën, Ilya de Groot, and Nienke Aaftink for their effort in data preparation.

The authors are grateful to SURFsara for the processing time on the Dutch national supercomputer (\url{www.surfsara.nl/systems/cartesius}). We gratefully acknowledge the support of NVIDIA Corporation with the donation of the Titan V GPU used for this research. 

E.E. Bron acknowledges support from Dutch Heart Foundation (PPP Allowance, 2018B011) and the Netherlands CardioVascular Research Initiative (Heart-Brain Connection: CVON2012-06, CVON2018-28). E.E. Bron and W.J. Niessen are supported by Medical Delta Diagnostics 3.0: Dementia and Stroke. V. Venkatraghavan and W.J. Niessen acknowledge funding from the Health~Holland LSH-TKI project Beyond (LSHM18049). This work is part of the EuroPOND initiative, which is funded by the European Union's Horizon 2020 research and innovation programme under grant agreement No. 666992.

The work described in this study was carried out in the context of the Health-RI Parelsnoer Neurodegenerative Diseases Biobank. Parelsnoer biobanks are part of and funded by the Dutch Federation of University Medical Centers and has received initial funding from the Dutch Government (2007-2011).

Data collection and sharing was funded by the Alzheimer's Disease Neuroimaging Initiative (ADNI) (National Institutes of Health Grant U01 AG024904) and DOD ADNI (Department of Defense award number W81XWH-12-2-0012). ADNI is funded by the National Institute on Aging, the National Institute of Biomedical Imaging and Bioengineering, and through generous contributions from the following: AbbVie, Alzheimer’s Association; Alzheimer’s Drug Discovery Foundation; Araclon Biotech; BioClinica, Inc.; Biogen; Bristol-Myers Squibb Company; CereSpir, Inc.; Cogstate; Eisai Inc.; Elan Pharmaceuticals, Inc.; Eli Lilly and Company; EuroImmun; F. Hoffmann-La Roche Ltd and its affiliated company Genentech, Inc.; Fujirebio; GE Healthcare; IXICO Ltd.; Janssen
Alzheimer Immunotherapy Research \& Development, LLC.; Johnson \& Johnson
Pharmaceutical Research \& Development LLC.; Lumosity; Lundbeck; Merck \& Co., Inc.; Meso Scale Diagnostics, LLC.; NeuroRx Research; Neurotrack Technologies; Novartis Pharmaceuticals Corporation; Pfizer Inc.; Piramal Imaging; Servier; Takeda Pharmaceutical Company; and Transition Therapeutics. The Canadian Institutes of Health Research is providing funds to support ADNI clinical sites in Canada. Private sector contributions are facilitated by the Foundation for the National Institutes of Health (www.fnih.org). The grantee organization is the Northern California Institute for Research and Education, and the study is coordinated by the Alzheimer’s Therapeutic Research Institute at the University of Southern California. ADNI data are disseminated by the Laboratory for Neuro Imaging at the University of Southern California.


\bibliographystyle{model2-names}\biboptions{authoryear}
\bibliography{adniparelsnoer.bib}

\clearpage
\beginsupplement
\section*{Supplementary files}

\begin{figure}[h]
\centering
\begin{subfigure}[b]{1\textwidth}
   \includegraphics[width=1\linewidth]{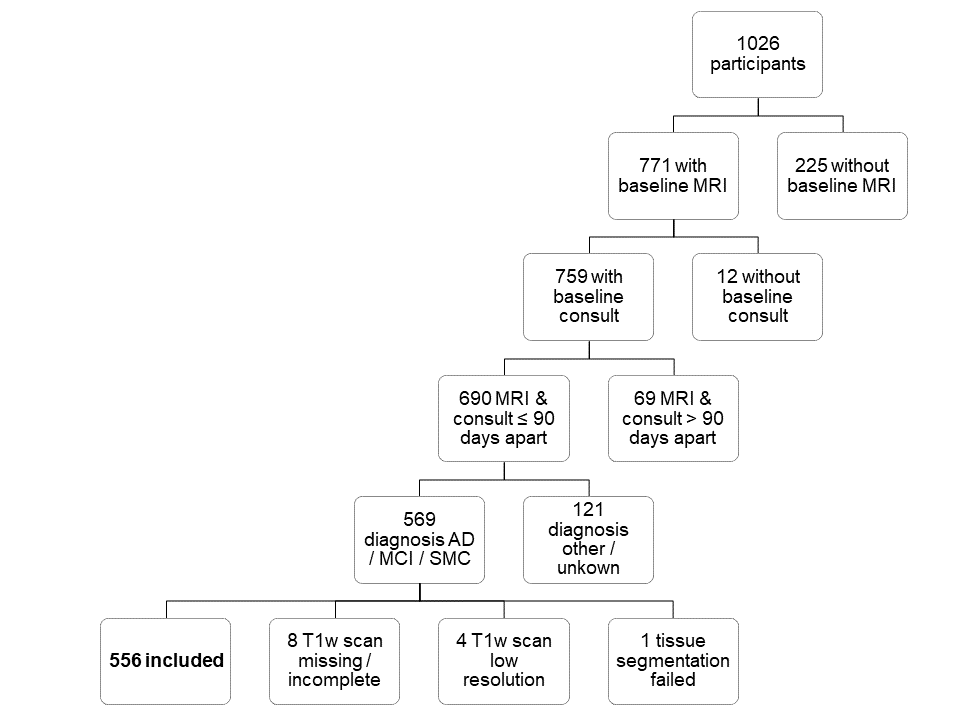}
\end{subfigure}

\caption[pnd]{Inclusion of participants of the PND data set}
\label{fig:pnd-inclusion}
\end{figure}

\end{document}